\documentclass[12pt]{article}

\usepackage{epsfig}

\begin{document}

\newcommand{\beq}{\begin{equation}}
\newcommand{\eeq}{\end{equation}}
\newcommand{\beqa}{\begin{eqnarray}}
\newcommand{\eeqa}{\end{eqnarray}}

\def\ov{\overline}
\def\onlyif{\rightarrow}

\def\openone{\leavevmode\hbox{\small1\kern-3.8pt\normalsize1}}

\def\a{\alpha}
\def\b{\beta}
\def\g{\gamma}
\def\r{\rho}
\def\minus{\,-\,}
\def\eks{\bf x}
\def\kay{\bf k}

\def\ket#1{|\,#1\,\rangle}
\def\bra#1{\langle\, #1\,|}
\def\braket#1#2{\langle\, #1\,|\,#2\,\rangle}
\def\proj#1#2{\ket{#1}\bra{#2}}
\def\expect#1{\langle\, #1\, \rangle}
\def\trialexpect#1{\expect#1_{\rm trial}}
\def\ensemblexpect#1{\expect#1_{\rm ensemble}}
\def\kpsi{\ket{\psi}}
\def\kphi{\ket{\phi}}
\def\bpsi{\bra{\psi}}
\def\bphi{\bra{\phi}}

\def\ditto{\rule[0.5ex]{2cm}{.4pt}\enspace}
\def\th{\thinspace}
\def\ni{\noindent}
\def\thirty{\hbox to \hsize{\hfill\rule[5pt]{2.5cm}{0.5pt}\hfill}}

\def\set#1{\{ #1\}}
\def\setbuilder#1#2{\{ #1:\; #2\}}
\def\Prob#1{{\rm Prob}(#1)}
\def\pair#1#2{\langle #1,#2\rangle}
\def\Id{\bf 1}

\def\dee#1#2{\frac{\partial #1}{\partial #2}}
\def\deetwo#1#2{\frac{\partial\,^2 #1}{\partial #2^2}}
\def\deethree#1#2{\frac{\partial\,^3 #1}{\partial #2^3}}

\newcommand{\xx}{{\scriptstyle -}\hspace{-.5pt}x}
\newcommand{\yy}{{\scriptstyle -}\hspace{-.5pt}y}
\newcommand{\zz}{{\scriptstyle -}\hspace{-.5pt}z}
\newcommand{\kk}{{\scriptstyle -}\hspace{-.5pt}k}
\newcommand{\sx}{{\scriptscriptstyle -}\hspace{-.5pt}x}
\newcommand{\sy}{{\scriptscriptstyle -}\hspace{-.5pt}y}
\newcommand{\sz}{{\scriptscriptstyle -}\hspace{-.5pt}z}
\newcommand{\sk}{{\scriptscriptstyle -}\hspace{-.5pt}k}

\def\openone{\leavevmode\hbox{\small1\kern-3.8pt\normalsize1}}

\title{A modular eballot system -- V0.6}
\author{Andrea Pasquinucci
\\
\small
{\it {\rm UCCI.IT}, via Olmo 26, I-23888 Rovagnate (LC), Italy
}}
\date{December 22, 2006}
\maketitle

\abstract{We consider a reasonably simple voting system which can be 
implemented for web-based ballots. Simplicity, modularity and the 
requirement of compatibility with current web browsers leads to a system 
which satisfies a set of security requirements for a ballot system which 
is not complete but sufficient in many cases. Due to weak-eligibility 
and vote-selling, this system cannot be used for political or similar 
ballots.}

\vspace{1 cm} 
\normalsize

\section{Introduction}

Most of current (commercial) web-based voting, balloting or polling 
system include little if no real security, in particular concerning 
privacy of the voter. Ballots are collected on a web site and stored in 
a database. Privacy of the voter is often guaranteed only as far as the 
vote is recorded in the database independently of the voting 
credentials. Indeed some commercial services have warnings that the 
system administrators can learn what every voter has voted for.

Here we present a simple, modular protocol, based on sound and common 
cryptography and previous results in the literature \cite{other,bruschi} 
which can be implemented as a web service compatible with current web 
browsers and not requiring any user education. Most of the cryptographic 
operations required in the protocol can be implemented with common tools 
like PGP or gnupg \cite{pgp}.

Our system does not satisfy all requirements for a complete voting 
system, for example it is possible to sell a vote, and for this reason 
it is not suitable for political election and in general in all cases in 
which are requested some of the properties that our system does not 
satisfies. For this reason we prefer not to call it a {\sl voting 
system} but only a {\sl balloting system}. We do not call it either a 
{\sl polling system} in which results are statistical, since our 
protocol guarantees that each vote is counted correctly and each 
eligible voter can cast her ballot.

In section 3 we list which requirements are satisfied by our 
system and which are not. In section 4 we describe our protocol, in 
section 5 we discuss how the security requirements are satisfied and in 
section 6 we discuss some threats and countermeasures.

In Appendix A we describe a variation of the protocol which uses blind 
signatures \cite{blindsig,bruschi}.

In a companion paper \cite{modultech}, we discuss how to implement 
our protocol in practice in its simplest version.

\section{Preliminaries}

We start by indicating the components of our protocol. There are 6 
human roles, who have well precise duties. As we will see, some aspects 
of the security of the protocol are based on the {\sl separation of 
duties} among the managers of the system. The 6 roles are:

\begin{enumerate}

\item {\sl the Voter}: no request is made on the Voter, except to be 
able to use a common web browser and in some cases to install a piece 
software on her own machine

\item {\sl the Authentication Managers (AuthMgr)} must organize the 
ballot selecting the eligible voters and providing them with the voting 
credentials; at the end of the voting period they must check that only 
eligible voters have casted a ballot

\item {\sl the Managers of the Authentication Web Server (AuthSysMgr)} 
must setup and run the Authentication Web Server and provide the AuthMgr 
with the list of voters that have casted a ballot at the end of the 
voting period

\item {\sl the Managers of the Anonymizer System (AnonSysMgr)} must 
guarantee that all connections to the Vote Web Server do not leak 
information on the IP address of the voter computer

\item {\sl the Vote Managers (VoteMgr)} must count the votes at the end 
of the voting period and announce the result of the ballot

\item {\sl the Managers of the Vote Web Server (VoteSysMgr)} must setup 
and run the Vote Web Server and provide, at the end of the voting 
period, the VoteMgr with the encrypted votes that have been casted.

\end{enumerate}

Moreover, the system is divided in 4 modules

\begin{enumerate}

\item {\sl the Voter (or Client)}: this is the voter machine and 
software, for example a personal computer with a web browser and in case 
a local specialized proxy

\item {\sl the Authentication Server (AuthSrv)} verifies the credentials 
of each voter and gives to the voter a Authentication Token (called 
VoteAuthorization) which allows the voter to cast her ballot

\item {\sl the Anonymizer} renders all accesses to the Vote Server 
anonymous

\item {\sl the Vote Server (VoteSrv)} allows the voter to cast only once 
the ballot in an anonymous way.

\end{enumerate}

\section{Requirements}

Some requirements often used in defining an electronic voting system 
are:

\begin{enumerate}

\item {\sl Unreusability (prevent double voting)}: no voter can vote 
twice

\item {\sl Privacy}: nobody can get any information about the voter's 
vote

\item {\sl Completeness}: all valid votes should be counted correctly

\item {\sl Soundness}: any invalid vote should not be counted

\item {\sl Individual verifiability}: each eligible voter can verify 
that her vote was counted correctly

\item {\sl Weak Eligibility}: only eligible voters can get voting 
credentials from trusted authorities

\item {\sl Eligibility}: no one who is not allowed to vote can vote

\item {\sl Fairness}: nothing can affect the voting

\item {\sl Robustness}: the voting system should be successful regarding 
of partial failure of the system

\item {\sl Universal Verifiability}: anyone can verify the fact that the 
election is fair and the published tally is correctly computed from the 
ballots that were correctly casted

\item {\sl Incoercibility}: a voter cannot be coerced into casting a 
particular vote by a coercer

\item {\sl Receipt-freeness}: a voter neither obtains nor is able to 
construct a receipt proving the content of her vote.

\end{enumerate}

Our system satisfies only the first 6 requirements. In particular it 
does not satisfies Eligibility (only Weak Eligibility) nor 
Receipt-freeness, both of which mean that a voter is able to prove to 
someone else how she has voted, and so to sell her vote.

\section{The modular protocol}

We describe now at high level the protocol and its components following 
the procedure of a ballot.

\subsection{Step 1 -- Distribution of credentials}

This step is done off-line. The Authentication Managers provide the 
voters with the voting credentials:

\begin{enumerate}

\item {\sl a username+password and/or SSL/TLS client certificate} which 
can be reused or shared among various voters and which allows the voters 
to be authenticated at the AuthSrv

\item {\sl a VoteToken}\footnote{This is also called a {\sl Secret Token} 
to stress the role of this credential.} (a PseudoRandom-string) unique for each voter 
and for each ballot, which can be used only once

\item {\sl the SSL/TLS fingerprints} of the web SSL/TLS certificates of 
the AuthSrv and VoteSrv.

\end{enumerate}

For higher security the username+password (and/or client certificate) 
and VoteToken should be distributed using different communication 
channels.

\subsection{Step 2 -- the Voter}

In principle the Voter can access the balloting system with {\sl one} of 
the following systems listed in increasing security. In the practical 
implementation of the protocol only one of the following must be used by 
all voters in the same ballot, that is before the ballot it must be 
decided which one of them the voters must use. The possible voter's 
systems are:

\begin{itemize}

\item {\sl a standard SSL/TLS web browser}

\item {\sl a standard web browser and a client proxy}: the client 
proxy is an application which runs on the voter's machine, receives 
http requests from the voter web browser, anonymizes the requests and 
connects securely to the AuthSrv and VoteSrv

\item {\sl a purpose built https client}: in practice a reduced 
version of a SSL/TLS web browser dedicated only to this purpose which 
includes the features both of the web browser and of the client proxy

\item {\sl a purpose built hardware device}: (for example something 
similar to a smartphone or handheld device) in this case the hardware, 
the operating system and the browser are custom made. 

\end{itemize}

\subsection{Step 3 -- Authentication Server}

The AuthSrv is a secure web server running an application able to 
implement the protocol which will be presented below. The AuthSrv, 
managed by the AuthSysMgr, needs to have:

\begin{enumerate}

\item {\sl the public key of the AuthMgr}

\item {\sl its own private/public key certificate} used to establish the 
SSL/TLS session with the voter's browser

\item {\sl its private/public key} to digitally sign the 
VoteAuthorizations (see below)

\item {\sl the public key of the VoteSrv} to encrypt the 
VoteAuthorizations

\item {\sl the list of username+password and/or client certificates}

\item {\sl the list of valid VoteTokens}.

\end{enumerate}

The authentication procedure is the following:

\begin{enumerate}

\item the voter connects to the AuthSrv with her web browser, checks the 
fingerprints of the server digital certificate and is authenticated 
using username+password and/or client digital certificate; finally the 
voter submits her VoteToken to the AuthSrv

\item the AuthSrv checks that the VoteToken has not been used and 
creates a unique PRN (Pseudo-Random-Number) called the {\sl 
VoteAuthorization} and optionally a short PIN (see below)

\item the AuthSrv digitally signs and encrypts with the VoteSrv public 
key the VoteAuthorization (if the PIN is used, this is added to the 
VoteAuthorization before encryption)

\item the AuthSrv records that the VoteToken has been used (for example 
by signing and encrypting it together with the current time and username 
with the AuthMgr public key and writing it to a file with the same 
name)

\item the AuthSrv sends to the voter's web browser the signed and 
encrypted VoteAuthorization. If the PIN is used, it is also sent to the 
voter not encrypted; the PIN must not be stored by the client-proxy nor 
forwarded automatically in clear to the VoteSrv, but only shown by the 
browser to the voter who should record it off-line; notice that the PIN 
is also included in the encrypted VoteAuthorization.

\end{enumerate}

The PIN could be useful if the vote phase (step 5 below) is done at a 
later time than the authorization phase (this step 3). In this case 
the voter must store the VoteAuthorization somewhere: the voter can 
write the VoteAuthorization on a piece of paper or, more likely, store 
it as a file on the computer either directly or in the client-proxy. In 
this second case there is the risk that someone else can access the 
voter computer, find the VoteAuthorization and use it. If a PIN is used 
and the voter has saved off-line the short PIN when has received the 
VoteAuthorization, then a third person will not be able to use the 
stolen VoteAuthorization.

\subsection{Step 4 -- Anonymizer}

The VoteSrv must not know nor be able to discover who is voting. Notice 
that if the voter uses a normal web browser some information, like the 
contents of the UserAgent field, are sent to the VoteSrv.

Since the connection between the voter and the VoteSrv is encrypted 
using SSL/TLS, the contents can be anonymized only by the client itself 
for example by the client-proxy.

But also the IP packets reaching the VoteSrv must be anonymized so not 
to disclose the IP address of the voter machine. This can be done in 
various ways of which we list some here in increasing security:

\begin{enumerate}

\item a NAT or double-NAT device: in this way all packets will reach the 
VoteSrv with the same source address

\item a Crowd  scheme \cite{crowd} 

\item a Chaum mixer \cite{chaum}

\item Tor \cite{tor} or similar scheme \cite{torpub}.

\end{enumerate}

The Manager of the Anonymizer must choose a anonymizer system in 
relation to the requirements of the ballot. Notice that some anonymizing 
schemes can introduce very large delays in the traveling of the IP 
packets, making it almost impossible to vote in real time, but giving 
higher guarantees of truly anonymize the traffic.

\subsection{Step 5 -- Vote Server}

The VoteSrv is a secure web server with an application able to implement
the protocol which will be presented below. The VoteSrv, managed by the
VoteSysMgr, needs to have

\begin{enumerate}

\item {\sl the public key of the AuthMgr}

\item {\sl the public key of the VoteMgr}

\item {\sl its own private/public key certificate} used to establish the
SSL/TLS session with the voter's web browser

\item {\sl its private/public key} to sign the Votes (see
below).

\end{enumerate}

The procedure to vote is the following:

\begin{enumerate}

\item the voter connects with her browser to the VoteSrv through the 
anonymizer and checks the fingerprints of the digital certificate of the 
VoteSrv used for the SSL/TLS session

\item the VoteSrv sends the Web Form to cast the ballot to the voter (if 
the PIN is required, this is requested in the Form)

\item the voter casts her ballot and sends to the VoteSrv the Form 
filled-in and the VoteAuthorization (this is usually done automatically 
by the browser or the client-proxy which has stored the 
VoteAuthorization for the voter)

\item the VoteSrv decrypts the VoteAuthorization and checks the 
signature of the AuthSrv, checks that the VoteAuthorization has not been 
already used and that the optional PIN given by the voter in the 
submitted Form matches the one inside the VoteAuthorization

\item the VoteSrv records that the VoteAuthorization has been used (for 
example by signing and encrypting it with the AuthMgr public key and 
writing it to a file with the same name)

\item the VoteSrv generates a digest of the vote including a precise 
timestamp and a random string as the {\sl VerificationCode}, signs and 
encrypts with the VoteMgr public key the vote together with the 
VerificationCode and writes it in a file with the same name (that is 
the name of the file is the VerificationCode)

\item the VoteSrv sends to the voter's web browser the VerificationCode, 
the digital signature of the VerificationCode, the precise timestamp and 
the random string. 
In this way the voter can recompute the VerificationCode and verify that 
it corresponds to her vote. The voter has to keep off-line the 
VerificationCode to check that her vote is correctly counted in the 
results of the ballot.

\end{enumerate}

\subsection{Step 6 -- Vote counting}

This step is done off-line and is composed by:

\begin{enumerate}

\item the AuthMgr receives from the AuthSysMgr the signed and encrypted 
VoteTokens, decrypts them, checks the signatures and records each
used VoteToken together with the time of use and the username which has 
used it

\item the AuthMgr receives from the AuthSysMgr the signed and encrypted
VoteAuthorizations created, decrypts them, checks the signatures and 
makes the list of {\sl created} VoteAuthorizations

\item the AuthMgr receives from the VoteSysMgr the signed and 
encrypted VoteAuthorizations used, decrypts them, checks the 
signatures and makes the list of {\sl used} VoteAuthorizations

\item the AuthMgr checks that the two lists of VoteAuthorizations are 
consistent (notice that according to the protocol there is no recorded 
information that links the VoteTokens to the VoteAuthorizations so that, 
unless the AuthSysMgr cheats by recording extra information and passing 
it to the AuthMgr, the list of used VoteTokens and used 
VoteAuthorizations are not linked)

\item the VoteMgr receives from the VoteSysMgr the signed and encrypted 
Votes, for the moment the VoteMgr does not decrypt them but makes the 
list of VerificationCodes (which are the file names)

\item the VoteMgr checks with the AuthMgr that the number of votes is 
consistent with the number of valid and used VoteAuthorizations (notice 
that according to the protocol there is no recorded information that 
links the Votes to the VoteAuthorizations so that, unless the VoteSysMgr 
cheats by recording extra information and passing it to the VoteMgr, the 
list of Votes and used VoteAuthorizations are not linked)

\item the AuthMgr publishes the list of used VoteTokens and the 
timestamp of their use so that voters can check that the time at which 
they obtained the VoteAuthorization is correct (this is the time
at which the voters have authenticated, not the time of vote); 
the AuthMgr publishes 
also the list of VoteTokens which have not been used, that is of 
eligible voters who did not vote

\item the VoteMgr publishes the list of VerificationCodes so that 
voters can check if their VerificationCode is present in the list; if a 
voter does not find her VerificationCode in the list, she must inform 
the VoteMgr of this before the votes are decrypted, in this way if 
there has been a problem in the procedure, the ballot can be canceled 
before counting the votes (this step also prevents a voter to false 
claim that her VerificationCode is not present if the ballot had a 
result she did not like); to report a missing VerificationCode, a voter 
should present to the VoteMgr her VerificationCode with its digital 
signature

\item after the VerificationCodes have been checked by the voters, the 
VoteMgr decrypts the signed and encrypted Votes, checks the signatures, 
makes the list of Votes and counts them

\item the VoteMgr publishes the list of VerificationCodes and 
individual Votes so that voters can check that their votes have been 
counted correctly, and publishes the result of the ballot.

\end{enumerate}

\section{Security requirements analysis}

We now consider the security properties of the protocol just described. 

In the following discussion when we say that a system manager cheats by 
modifying the software so that it does not follow the protocol 
described, it is equivalent to the fact that the server has fallen under 
control of an attacker who can then do what a system manager can do.

\begin{description}

\item[Unreusability:] both the VoteToken and the VoteAuthorization can be 
used only once, at the AuthSrv and the VoteSrv respectively; the 
AuthSysMgr could issue more VoteAuthorization for the same VoteToken, 
analogously the VoteSysMgr could allow multiple votes for the same 
VoteAuthorization, but this is discovered by the cross checking of the 
number of VoteTokens, VoteAuthorizations and votes done by the AuthMgr 
and VoteMgr.

\item[Privacy:] the information received by the VoteSrv does not allow 
the VoteSysMgr to learn who is casting a vote; privacy can be 
theoretically violated only if at least two system managers are 
cheating:

\begin{itemize}
  
\item if the AuthSysMgr records the connection between the the VoteToken 
and the VoteAuthorization and the VoteSysMgr records the connection 
between the VoteAuthorization and the Vote, then by combining this 
information is possible to know who has voted what

\item if the AnonSysMgr records the IP addresses (and tcp ports) of the 
voter machine when connecting to the VoteSrv and the time of connection, 
and the VoteSysMgr records the connection between the VoteAuthorization 
and the Vote and the time of the vote, then by combining this 
information is possible to know which client IP has casted a vote and 
at what time.

\end{itemize}

\item[Completeness:] it is not possible to modify a vote and all votes 
are counted, so that it is not possible to force to count the votes in a 
wrong way.

\item[Soundness:] every vote is signed by the AuthSrv and encrypted with 
the VoteSysMgr public key so that it is not possible to count invalid 
votes.

\item[Individual verifiability:] a verification code is given to each 
voter to allow to verify that her vote has been counted correctly.

\item[Weak Eligibility:] the AuthSrv verifies that only eligible voters 
can obtain a VoteAuthorization; the VoteSrv checks that the 
VoteAuthorization has been issued by the AuthSrv but not directly the 
voter credentials, thus obtaining the weak eligibility requirement; the 
AuthMgr could issue VoteAuthorizations to non eligible voters or the 
VoteMgr could accept non valid VoteAuthorizations, but the vote counting 
procedure by the AuthMgr and VoteMgr and the checking of the public 
results by the voters will discover it.

\end{description}

\section{Threats analysis}

Although the protocol guarantees its self-correctness, in practice two 
factors can reduce its security:

\begin{enumerate}

\item implementation errors (like software bugs)

\item human intervention.

\end{enumerate}

We analyze threats from two points of view: the purpose of the attack 
and the system attacked.

\subsection{Threat model}

An attacker can be interested to 

\begin{enumerate}

\item influence the final result (system integrity threat)

\item capture information on votes (privacy threat)

\end{enumerate}

\subsubsection{System Integrity}

\noindent System Integrity attacks can be performed by:

\begin{enumerate}

\item multiple use of credentials

\item vote modification

\begin{itemize}

\item by the VoteSrv

\item by network man-in-the-middle attacks

\item on the voter's machine

\end{itemize}

\item stealing of vote credentials from the AuthSrv.

\end{enumerate}

All of these threats, except for that on the voter's machine, are 
mitigated by countermeasures in the protocol so that any such attempt 
will be discovered.

The protocol does not have countermeasures against {\sl Vote Selling}, 
indeed it lacks the {\sl receipt-freeness} property since it bases some 
of its security properties on the receipt given to the voter by the 
VoteSrv. Indeed a voter can prove to a third party, if she wishes, how 
she has voted by showing the VerificationCode, or a voter can pass her 
voting credentials or VoteAuthorization to someone else and have the 
attacker cast directly the vote.

\subsubsection{Privacy}

\noindent Privacy attacks can be performed

\begin{enumerate}

\item by sniffing on the connections between the voter and the servers: 
this threat is countered by encrypting all traffic

\item by installing sniffers on the AuthSrv and VoteSrv to record all 
information: this threat must be countered by the system managers

\item if at least two system managers collude in violating the system

\item by installing a sniffer on the voter machine: the protocol does 
not provide any countermeasure to this threat.

\end{enumerate}

\subsection{System vulnerabilities}

Attacks by intruders in a system can lead to complete compromise of 
the machine so that the attackers takes complete control of it, in 
other words they are able to act as the system manager. Notice that we 
need to trust the system managers and the correct status of the systems 
for the protocol to work as stated. 

Compromise of the voter machine can lead to the loss of privacy and the 
possibility for the attacker to modify the vote (loss of integrity). 
Voter are able to recompute their VerificationCodes, and they can 
discover if their votes have been modified. (Notice that today this kind 
of attack is the most likely and easy to implement.)

Compromise of the AuthSrv can lead to the stealing of vote credentials 
and loss of privacy if there is also a compromise of the VoteSrv. These 
risks are reduced due to countermeasures in the protocol.

Compromise of the Anonymizer can lead to a loss of privacy if there is 
also a compromise of the VoteSrv.

Compromise of the VoteSrv can lead to modification of votes, 
countermeasures for this risks are present in the protocol. Compromise 
of the VoteSrv together with the compromise of the Anonymizer or AuthSrv 
can also lead to a loss of privacy.

\subsection{Off-line threats}

The AuthMgr and VoteMgr must be trusted. If they cheat they can modify 
the results of the ballot. In particular:

\begin{itemize}

\item the AuthMgr can distribute vote credentials to not eligible voters 
(no countermeasures)

\item the AuthMgr can give more than one vote credential to the same 
voter (no countermeasures)

\item the VoteMgr could fabricate votes and add them to the final tally, 
but this is discovered by the AuthMgr checking the number of votes 
w.r.t.\ the number of voters and the voters verifying that their votes 
have been counted correctly

\item the VoteMgr can change the votes of the voters, but this is 
discovered by the individual checks of the voters.

\end{itemize}

\subsection{Other threats}

\subsubsection{Attacks to the VerificationCode}

Format of the VerificationCode is crucial to prevent the VoteSysMgr to 
modify votes. Assume for example that the VerificationCode is a PRN. In 
this case the VoteSysMgr could modify the protocol as follows. Suppose 
there is a Yes/No ballot and the VoteSysMgr wants to fix the result for 
No. The first voter that votes Yes is processed correctly, all other 
voters who vote Yes (or enough of them to make No win) will have their 
vote changed to No but will receive the same VerificationCode of the 
first voter. In this way all Yes voters will have the same 
VerificationCode corresponding to a published and correctly counted Yes 
vote. Only by checking the VerificationCode among Yes voters it would be 
possible to discover that they are all equal, which must not be since 
they should be PRN. But voters must keep confidential their 
VerificationCodes since otherwise the privacy of their votes will be 
lost.

For this reason the VerificationCode is a digest of the vote and of the 
time of the vote, in such a way that the voter can recompute the digest 
using as ingredients her vote and the timestamp.

But a public computable VerificationCode which includes {\sl only} 
the vote and a timestamp is liable to a brute force attack. 
Indeed the vote and the procedure to compute the VerificationCode are 
public and the only missing information is the timestamp. By computing 
all VerificationCodes with that vote and all reasonable timestamps, 
the attacker can find the correct timestamp. By matching the computed 
timestamp to the time of the authentication recorded by the AuthSrv and 
published by the AuthMgr, the attacker could, in some circumstances, 
discover the name of the voter.

To prevent this brute force attack, a sufficiently long random string 
must be added to the computation of the VerificationCode. Since the 
VerificationCode is nothing else than a cryptographical hash of the vote, 
the precise timestamp and the random string, and the voter is asked to
verify that the timestamp is correct, it is not possible for the VoteSysMgr 
to mount the attack discussed at the beginning of this section. 
Indeed if it was possible, it would mean that by adjusting the random string
the VoteSysMgr would be able to construct two different messages 
(they are different due to the different timestamps) with the same 
cryptographical hash. But this is not possible by definition of a 
cryptographical hash function.

\subsubsection{Human verifiability}

In this protocol a large part of the security relies on the behavior of 
the people involved, both the managers and officials of the ballot and 
the voters themselves. In particular the voters must check that their 
votes are counted correctly, using their VerificationCode, to prevent 
various possible attacks. It is reasonable to think that most voters who 
have voted will check their vote. It is less likely that voters will 
recompute their VerificationCode since, even if it is a very simple 
procedure, this requires a minimum knowledge about cryptography. Thus 
the re-computation of the VerificationCode is something that will be done 
by experts only. Anyway, the fact that some voters will do it, will be 
able, at least statistically, to prevent some attacks.

One possible attack venue is to consider voters who do not vote. In each 
ballot there are always voters that do not cast their vote for many 
different reasons. It will be very difficult to convince these voters to 
check that they correctly appear as not having voted in the final lists 
published by the AuthMgr. Assuming that all eligible voters who do not 
vote do not check this list, the following attack is possible. The 
AuthSysMgr logs in to the AuthSrv a few minutes before the end of the 
ballot and gets the list of not used VoteTokens.\footnote{Actually on the 
AuthSrv there are only the Hash of the VoteTokens, so the AuthSysMgr must 
also have obtained, for example from the AuthMgr, the original list of 
VoteTokens.} 
The AuthSysMgr must also have obtained, for example from the AuthMgr, 
the list of all username+passwords. With these data, 
the AuthSysMgr can vote instead of all the voters who have not voted and 
in this way violate the integrity of the ballot. It is still possible to 
discover this attack by checking the integrity of the AuthSrv since the 
AuthSrv must be sealed during the time of the ballot and all logins to 
it must be recorded in the logs. In a correct procedure, the AuthMgr 
(not the AuthSysMgr) should seal the machine before the beginning of the 
ballot and remove the seal at the end of the ballot but only after 
having checked that nobody has logged in the machine during the ballot 
and no breach of security have happened. (If the AuthSysMgr has 
installed a program which notifies him of the not used VoteTokens, then 
the machine is not secure to start with.) Thus it is possible to 
discover this attack even if all eligible voters who do not vote do not 
check the list of voters, but it requires a very careful implementation 
of the procedures, both human and automatic.

Always considering the voters who do not vote, since the authentication 
is automatically done by the AuthSrv, if a voter complains that she has 
not voted but her VoteToken appears in the list of the used ones, the 
AuthMgr has to understand what really has happened. There are a few 
possibilities and among them the following

\begin{itemize}

\item someone has stolen the voting credentials of the voter and, 
knowing that the voter would have not used them, has voted in her place

\item the voter has sold (or has been forced to sell) her voting 
credentials but now is trying to get the ballot cancelled

\item a violation of the AuthSrv has happened, as described before, and 
unsed voting credentials have been used by attackers

\item there has been an error in the procedures, for example in the 
listing and counting of votes and VoteTokens

\item the voter has in reality casted her ballot, but has decided to try 
to cancel the ballot without even knowing the result (in this case is 
the voter who is cheating).

\end{itemize}

It is up to the ballot officials to understand what has really 
happened and decide if the results of the ballot are correct or there 
has been some violations.

\section{Conclusions}

The effectiveness and security of the protocol are based on 

\begin{itemize}

\item trust in the people involved in the protocol: they must not 
collude against the protocol

\item absence of bugs in software which could render insecure the 
implementation of the protocol: from operating system to libraries and 
applications

\item systems correctly managed and in integral state: absence of worms, 
viruses, trojans, rootkits etc.\ on all systems involved, included the 
voter's machine.

\end{itemize}

Under these conditions we believe that our protocol can be practically 
implemented and can deliver a web based balloting system with much 
higher security than those currently commercially available.

\section*{Appendix A}

In ref.\ \cite{bruschi} was proposed a protocol similar to ours but 
based on blind signatures \cite{blindsig}. Our protocol can be easily 
modified to support blind signatures. 

The main implementational difference between our protocol and a version 
using blind signatures is that the VoteAuthorization is not created by 
the AuthSrv but by the voter's client proxy (blind signatures are not 
possible with current web browsers). 

The procedure in Step 3 is then modified as follows:

\begin{enumerate}

\item the voter connects to the AuthSrv with her web browser through the 
client proxy, checks the fingerprints of the server digital certificate 
and is authenticated using username+password and/or client digital 
certificate

\item the voter creates a PRN (not unique, so there is a very small 
chance that two voters will create the same PRN in which case only one 
of them will be able to vote; the probability of this depends on the 
space of the PRN w.r.t.\ the number of voters), prepares a blind 
signature of it and sends to the AuthSrv the blind signature and the 
VoteToken

\item the AuthSrv checks the VoteToken and does the blind signature, 
that is digitally signs the encrypted PRN without knowing the value of 
the PRN

\item the AuthSrv records that the VoteToken has been used (for example 
by signing and encrypting it together with the current time and username 
with the AuthMgr public key and writing it to a file with the same name)

\item the AuthSrv sends the blind signature to the voter who extracts 
the signed PRN which is her VoteAuthorization.

\end{enumerate}

In this case the AuthSrv does not know and cannot compute the 
VoteAuthorization, so even if the AuthSysMgr would collude with the 
VoteSysMgr, they will not be able to obtain information on which voter 
has which VoteAuthorization. On the other hand, if blind signatures are 
used there is the (possibly small) risk that two voters will create the 
same VoteAuthorization and one of them will be prevented from voting.

Afterwords in the protocol, in step 5, the client-proxy encrypts the 
VoteAuthorization with the public key of the VoteSrv before sending it. 
Thus the AuthMgr should provide the voter together with the voting 
credentials also with the public key of the VoteSrv.

To introduce blind signatures in our protocol, it is then sufficient for 
the voter to use a client proxy and to modify accordingly the AuthSrv. 
The rest of the protocol does not change.

The protocol with blind signatures has the advantage of eliminating the 
possibility of collusion between the AuthSysMgr and the VoteSysMgr to 
violate the privacy of the voters. On the other side, if blind 
signatures are used, it is not possible to guarantee that all eligible 
voters can vote. For this reason, the protocol proposed in ref.\ 
\cite{bruschi} was called a {\sl polling} protocol in which statistical, 
but not exact, results are obtained.

Notice also that blind signatures can not be implemented with current 
web browsers. Thus implementing practically the protocol with blind 
signatures today is technically more complicated because it requires 
designing, realizing and maintaining a program which should run on 
every voter's machine.

\section*{Acknowledgments} 

We thank D.\ Bruschi and A.\ Lanzi for discussions, and H.\ 
Bechmann-Pasquinucci for inspiring remarks.


\begin{thebibliography}{99}

\bibitem{other} For a recent review of electronic voting systems, 
mixnet etc., see B.\ Adida, {\sl Advances in cryptographic voting 
systems}, PhD thesis 2006, MIT; see also D.\ Gritzalis editor, {\sl 
Secure electronic voting}, Kluwer Academic Publisher 2002, and B.\ 
Schneier, {\sl Applied Cryptography}, Wiley 1996

\bibitem{bruschi} D.\ Bruschi, I.\ Nai Fovino, A.\ Lanzi, {\sl A 
Protocol for Anonymous and Accurate E-Polling}, Proceedings of {\sl 
E-Government: Towards Electronic Democracy}, International Conference 
TCGOV 2005, Bolzano Italy 2005, Lecture Notes in Computer Science 3416 
Springer 2005

\bibitem{pgp} see e.g.\ rfc1991, rfc2440, rfc3156, http://www.pgp.com/, 
http://www.gnupg.org/

\bibitem{blindsig} D.\ Chaum, {\sl Security without identification; 
transaction systems to make big brother obsolete}, Comm.\ ACM 28(19) 
1985, 1030

\bibitem{modultech} A.\ Pasquinucci, {\sl Implementing the modular 
eballot system V1.0}, 2006, http://eballot.ucci.it/

\bibitem{crowd} M.K.\ Reiter, A.D.\ Rubin, {\sl Anonymous web 
transactions with Crowds}, Comm.\ ACM 42(2) 1999, 32

\bibitem{chaum} D.\ Chaum, {\sl Untraceable electronic mail, return 
address and digital pseudonyms}, Comm.\ ACM 24(2) 1981, 84

\bibitem{tor} {\sl The Onion Router}, http://tor.eff.org/

\bibitem{torpub} R.\ Dingledine, N.\ Mathewson, P.\ Syverson, {\sl Tor: 
the second-generation onion router}, http://tor.eff.org/

\end{thebibliography}
\end{document}